\documentclass{article}
\usepackage[utf8]{inputenc} 
\usepackage[T1]{fontenc}    
\usepackage{hyperref}       
\usepackage{url}            
\usepackage{booktabs}       
\usepackage{amsfonts}       
\usepackage{nicefrac}       
\usepackage{microtype}      
\usepackage[utf8]{inputenc} 
\usepackage[T1]{fontenc}    
\usepackage{hyperref}       
\usepackage{url}            
\usepackage{booktabs}       
\usepackage{nicefrac}       
\usepackage{microtype}      
\usepackage{bbm,mathbbol}
\usepackage{yhmath}
\usepackage{textcomp}
\usepackage{algpseudocode}

\usepackage{enumerate}

\usepackage[english]{babel}
\usepackage{blindtext}
\usepackage{csquotes}
\usepackage{graphicx}
\usepackage{amsmath}	
\usepackage{xcolor}
\usepackage{colortbl}
\usepackage{amssymb}
\usepackage{array}
\usepackage{booktabs}
\newcolumntype{L}{>{\centering\arraybackslash}m{3cm}}
\usepackage{algorithm}
\usepackage{amsmath}
\usepackage[utf8]{inputenc}
\usepackage{pgfplots}
\DeclareUnicodeCharacter{2212}{−}
\usepgfplotslibrary{groupplots,dateplot}
\usetikzlibrary{backgrounds}
\usepackage{xspace}
\usepackage{multirow}
\usepackage{wrapfig}
\usepackage{enumitem}
\usepackage{placeins}
\usepackage{amsthm}
\usepackage{enumerate}

\usepackage[colorinlistoftodos,prependcaption,textsize=scriptsize]{todonotes}

\newcommand{\bb}{\ensuremath{\mbox{B\@B}}}

\newcommand{\jochen}[1]{\todo[color=green!100!black]{J:~#1}}

\newcommand{\arash}[1]{\todo[color=blue!60!gray!30]{A:~#1}}
\newcommand{\leian}[1]{\todo[color=blue!60!gray!30]{L:~#1}}

\usepackage[preprint]{amlc_guide}

\usepackage{pdfprivacy}

\usepackage[utf8]{inputenc} 
\usepackage[T1]{fontenc}    
\usepackage{hyperref}       
\usepackage{url}            
\usepackage{booktabs}       
\usepackage{amsfonts}       
\usepackage{nicefrac}       
\usepackage{microtype}      
\usepackage{subcaption}
\usepackage{graphicx}
\usepackage{comment}
\usepackage{adjustbox}
\usepackage{wrapfig}

\title{A Constraint-Aware Generative Framework \\for Synthetic Origin-Destination Demand \\in Logistics Networks}

\author{%
  Leian Chen\\
}

\begin{document}
\maketitle
\vspace{-15pt}
\begin{abstract}
 Large-scale logistics networks require synthetic data generation capabilities to support scenario-based planning under novel conditions—such as network reconfiguration and demand shocks. Existing approaches, which rely primarily on historical observations, lack the ability to generate demand patterns that adapt to changes in network topology while respecting operational constraints. We propose a constraint-aware conditional generative framework for synthetic origin–destination demand generation in hierarchical logistics networks. The framework models demand as a conditional distribution over destinations given each origin, enabling topology-aware synthesis that is both topologically realistic and operationally feasible. Operational guidance is incorporated directly into the generative objective via differentiable constraints, while a flexible conditioning mechanism supports various operational contexts and adaptation to evolving network configurations. We instantiate the proposed framework based on a conditional generative model. Experimental validation on industrial real fulfillment and transportation network demonstrates 16\% improvement over graph neural network baselines, 87\% operational compliance, and efficient cold-start adaptation, enabling applications in capacity planning, network design evaluation, and routing optimization.
\end{abstract}
\vspace{-10pt}

\section{Introduction}
\label{sec_1}
\vspace{-5pt}
As fulfillment and transportation network grows in scale and complexity, planning teams face increasing pressure to evaluate decisions under future, novel, or uncertain conditions. Critical planning scenarios include testing new sort center placements, planning for demand surges, and designing networks under different growth scenarios. However, existing approaches face fundamental limitations: historical data only reflects past patterns under existing conditions, new facilities lack relevant data entirely, and planners cannot simulate novel configurations or stress-test against edge conditions. Traditional demand forecasting provides only point estimates based on these historical patterns, while scenario-based planning requires diverse sets of plausible demand realizations. Existing methodologies further suffer from network adaptation failures (e.g., regionalization), optimal guidance ignoring real-world constraints (e.g., inventory positions, capacity bounds), and temporal inconsistency (e.g., lack of day-over-day consistency).

These constraints hinder the ability to proactively evaluate network designs, conduct robust scenario planning, or develop resilient models under varied future states. We focus on generating synthetic origin-destination (O-D) demand data—the foundation for virtually all downstream transportation and capacity planning decisions in logistics networks. Large-scale transportation network exemplifies the unique complexities of this problem: a hierarchical structure spanning fulfillment centers (FCs), sort centers (SCs), delivery stations (DS) and ZIP destinations with asymmetric flows, dynamic capacity constraints, and frequent topology changes as facilities open or close. These characteristics create three fundamental challenges: (1) \textbf{topological realism}—respecting directional flow preferences inherent in logistics networks (e.g., FC→SC→DS flows are preferred over reverse paths), where 38{,}000+ ZIP destinations create O-D matrices with millions of entries that are computationally intractable for direct modeling; (2) \textbf{operational feasibility}—integrating forward-looking guidance values and constraints (e.g., capacity bounds, regional service requirements) under diverse future conditions that differ from historical observations; and (3) \textbf{adaptability}—maintaining performance under topology changes (e.g., facility openings, closures) without complete retraining.

\textbf{Technical Challenges and Related Work:} Classical spatial interaction 
models~\cite{wilson1971family,simini2012universal,ortuzar2011modelling} assume symmetric 
flows and cannot enforce operational constraints. Deep learning approaches to O-D flow 
modeling~\cite{simini2021deep,shi2020predicting} improve upon classical methods but focus 
on urban mobility prediction without graph-structured generation or constraint mechanisms. 
Graph convolutional networks~\cite{kipf2017semi} aggregate neighborhood information 
uniformly without attention, while graph attention networks~\cite{velickovic2018graph} 
introduce learned attention but do not encode transportation-specific flow hierarchies. 
Graph generative models~\cite{you2018graphrnn,simonovsky2018graphvae} lack constraint 
satisfaction mechanisms. Differentiable optimization 
layers~\cite{donti2017task,amos2017optnet,agrawal2019differentiable} integrate constraints but address only convex settings without topology adaptation. These 
gaps—the absence of constraint-aware generation, transportation-specific graph encoding, 
and topology adaptation—motivate the framework proposed in this paper (extended discussion 
in Appendix~\ref{sec:related_work_extended}).

\textbf{Our Contributions:} To address the limitations of existing demand synthesis approaches, we propose a constraint-aware conditional generative framework for O-D demand generation in hierarchical logistics networks. Our contributions are threefold:
\vspace{-5pt}
\begin{enumerate}
\item \textbf{Problem Formulation}: We formulate synthetic O-D demand generation as a constraint-aware conditional generative problem, where demand distributions are conditioned on network topology, operational guidance and other contexts, enabling principled demand synthesis under previously unseen network configurations.
\item \textbf{Topology- and Constraint-Aware Generation}: We introduce a general mechanism for integrating network structure and operational feasibility requirements into the generative objective through differentiable constraints, allowing generated demand to satisfy both topological realism and operational guidance.
\item \textbf{Topology Adaptation via Knowledge Transfer}: We propose a selective transfer learning strategy that enables efficient adaptation of the generative process to evolving network topologies while preserving learned structural regularities.
\end{enumerate}
\vspace{-5pt}
We demonstrate one instantiation of the proposed framework using a conditional generative model with topology-aware and constraint-aware conditioning.  To the best of our knowledge, this work represents the first systematic framework for constraint-aware O-D demand generation in large-scale hierarchical logistics networks, with direct applicability to  industrial transportation operations.

The remainder of this paper is organized as follows: Section \ref{sec_2} formulates the problem and framework overview. Section \ref{sec_3} presents an instantiation of our constraint-aware generative framework including network structure encoding, conditional demand generator, constraint satisfaction mechanisms, and cold start adaptation. Section \ref{sec_4} describes experiments and analysis on an industrial transcontinental fulfillment network. Section \ref{sec_6} concludes the work. The Appendix provides detailed implementation and deployment considerations.
\vspace{-5pt}
\section{Problem Formulation and Generative Framework Overview}
\label{sec_2}
\vspace{-5pt}

These three fundamental challenges in Section \ref{sec_1} motivate formalizing the problem as learning a constraint-aware conditional generative model over hierarchical transportation graphs. Let $G = (V, E, \mathcal{T})$ represent the transportation network, where $V = V_{\text{FC}} \cup V_{\text{SC}} \cup V_{\text{DS}} \cup V_{\text{ZIP}}$ is the node set with fulfillment centers, sort centers, delivery stations and ZIP destinations, $E \subseteq V \times V$ is the edge set of feasible transportation lanes constructed from operational routing data with attributes encoding distance and transit time, and $\mathcal{T}: V \rightarrow \{\text{FC}, \text{SC}, \text{DS},\text{ZIP}\}$ assigns node types.

The model generates demand probability distributions over destinations for each origin rather than absolute demand values, enabling flexible scenario planning. Following generative modeling conventions, we introduce a latent representation $\mathbf{z}$ to enable diverse sample generation. The model is parameterized by $\theta \in \Theta$ and generates O-D demand probability distributions $\widehat{\mathbf{P}} \in \Delta^{|V_{\text{FC}}| \times K}$ conditioned on the transportation graph $G$, conditional features $\mathbf{c} \in \mathbb{R}^{|V_{\text{FC}}| \times d_c}$ (operational context such as demand intensity, network utilization, and package size mix), and latent representation $\mathbf{z}$, where $K \in \mathbb{N}$ denotes the number of destination clusters, $d_c \in \mathbb{N}$ is the conditional feature dimension, each row $\widehat{\mathbf{P}}_{i,:} \in \Delta^K$ represents the outbound demand distribution from FC $i$, and $\Delta^K = \{\mathbf{p} \in \mathbb{R}_+^K : \sum_{k=1}^K p_k = 1\}$ is the $K$-dimensional probability simplex.

This probability-based formulation directly addresses all three challenges: destination clustering reduces complexity from $O(|V_{\text{FC}}| \times |V_{\text{ZIP}}|)$ to $O(|V_{\text{FC}}| \times K)$, probability distributions decouple spatial patterns from absolute volumes for flexible scenario planning, and graph-conditioning supports topology adaptation.

\textbf{Learning (training) phase:} Let $\mathcal{D}$ denote the empirical distribution of observed training data $\{(\mathbf{P}^{(n)}, G^{(n)}, \mathbf{c}^{(n)})\}_{n=1}^N$ where $N \in \mathbb{N}$ is the number of training samples. Model parameters $\theta$ are learned to generate realistic O-D probability distributions while encouraging satisfaction of operational constraints through an expected loss that combines data fidelity and constraint penalties:
\vspace{-5pt}
\begin{align}
\min_{\theta} \;
\mathbb{E}_{(\mathbf{P}, G, \mathbf{c}) \sim \mathcal{D}}
\Big[
\mathcal{L}_{\text{data}}(\mathbf{P}, \widehat{\mathbf{P}})
+ \sum_{\ell=1}^{L} \lambda_{\ell} \, \mathcal{L}_{\ell}(\widehat{\mathbf{P}})
\Big],
\vspace{-13pt}
\label{eq:generic_training_objective}
\end{align}
where $\widehat{\mathbf{P}}$ is generated by the model conditioned on $(G, \mathbf{c}, \mathbf{z})$, $\mathbf{z}$ encodes stochastic variations in demand patterns, $\mathcal{L}_{\text{data}}: \Delta^{|V_{\text{FC}}| \times K} \times \Delta^{|V_{\text{FC}}| \times K} \rightarrow \mathbb{R}_+$ is a model-specific data fidelity term (e.g., reconstruction loss for Variational Autoencoders (VAE)~\cite{VAE}~\cite{cVAE} or adversarial loss for Generative Adversarial Networks (GAN)~\cite{GAN}~\cite{cGAN}), $\{\mathcal{L}_{\ell}: \Delta^{|V_{\text{FC}}| \times K} \rightarrow \mathbb{R}_+\}_{\ell=1}^{L}$ are $L \in \mathbb{N}$ configurable constraint violation measures, and $\{\lambda_{\ell} \in \mathbb{R}_+\}_{\ell=1}^{L}$ control the relative importance of each constraint.

\textbf{Generation (inference) phase:} Synthetic constraint-aware probability distributions $\widehat{\mathbf{P}}$ are obtained by generating from the trained model parameterized by $\theta$ with latent codes $\mathbf{z} \sim p(\mathbf{z})$ and conditional inputs $(G, \mathbf{c})$, and an optional constraint facilitation:
\vspace{-2pt}
\begin{equation}
\widehat{\mathbf{P}} = \text{Generator}_\theta(G, \mathbf{c}, \mathbf{z}),
\qquad
\widetilde{\mathbf{P}} = \Pi(\widehat{\mathbf{P}}),
\end{equation}
where $\Pi: \Delta^{|V_{\text{FC}}| \times K} \rightarrow \Delta^{|V_{\text{FC}}| \times K}$ denotes an optional constraint facilitation operator (e.g., filtering or rescaling), and $\widetilde{\mathbf{P}} \in \Delta^{|V_{\text{FC}}| \times K}$ is the final feasible probability matrix. These cluster-level probabilities are then converted to operational demand through volume allocation: $D_{i,k} = V_i \cdot \widetilde{P}_{i,k}$, where $V_i \in \mathbb{R}_+$ represents the total demand volume allocated to facility $i$, and $\widetilde{P}_{i,k} \in [0,1]$ are the constraint-compliant probability values with $\sum_{k=1}^K \widetilde{P}_{i,k} = 1$. The framework operates on this compressed cluster space for computational tractability, with cluster-level demand subsequently expanded to full ZIP-level resolution through volume-weighted mapping. This two-stage design decouples model learning from optional constraint enforcement and destination expansion, enabling the framework to support a wide range of generative architectures—including VAEs, GANs, and other neural generators—without modification to the core formulation.

\vspace{-5pt}
\section{Instantiating the Constraint-Aware Generative Framework}
\label{sec_3}
\vspace{-5pt}

We instantiate the model-agnostic framework from Section~\ref{sec_2} using a conditional VAE (cVAE) architecture. The framework supports alternative generative models (e.g., GANs) through the same constraint satisfaction and transfer learning mechanisms. The instantiation consists of: (1) network structure encoding via graph attention (GAT), (2) conditional demand generator via cVAE, (3) constraint satisfaction mechanism, and (4) selective transfer learning for adaptation.
\vspace{-5pt}

\subsection{Network Structure Encoding}
\vspace{-5pt}
\label{subsec:GAT_architecture}
To encode the network topology $G$ into node representations that respect logistics hierarchies, we instantiate with a transportation-aware graph attention (GAT) mechanism. Standard GAT computes uniform attention across edge types, misrepresenting logistics networks where certain flows (e.g., FC→SC→DS) are preferred while others (e.g., DS→FC) are rare.

To address this limitation, we develop a transportation-aware attention mechanism that incorporates domain knowledge about logistics hierarchies through biased edge attributes. Our approach applies configurable flow bias parameters $\beta_{s,t}$ to edge attributes before standard GAT attention computation:
\begin{equation}
\mathbf{e}_{ij}^{\text{biased}} = \beta_{\mathcal{T}(i),\mathcal{T}(j)} \cdot \mathbf{e}_{ij},
\vspace{-1pt}
\end{equation}
where $\mathbf{e}_{ij} \in \mathbb{R}^{d_e}$ are the original edge attributes with $d_e \in \mathbb{N}$ the edge feature dimension, and $\beta_{s,t} \in \mathbb{R}_+$ are transportation-specific bias parameters for node types $s, t \in \{\text{FC}, \text{SC}, \text{DS}, \text{ZIP}\}$. The standard GAT attention coefficient $\alpha_{ij}^{(l)}$ for edge $(i,j)$ at layer $l$ then operates on these biased edge attributes:
\begin{equation}
\alpha_{ij}^{(l)} = \frac{\exp(\text{LeakyReLU}(\mathbf{a}^T[\mathbf{W}^{(l)}\mathbf{h}_i^{(l)} || \mathbf{W}^{(l)}\mathbf{h}_j^{(l)} || \mathbf{e}_{ij}^{\text{biased}}]))}{\sum_{k \in \mathcal{N}(i)} \exp(\text{LeakyReLU}(\mathbf{a}^T[\mathbf{W}^{(l)}\mathbf{h}_i^{(l)} || \mathbf{W}^{(l)}\mathbf{h}_k^{(l)} || \mathbf{e}_{ik}^{\text{biased}}]))},
\vspace{-3pt}
\end{equation}
where $\mathbf{h}_i^{(l)} \in \mathbb{R}^{d^{(l)}}$ is the representation of node $i$ at layer $l$, $d^{(l)} \in \mathbb{N}$ is the hidden dimension at layer $l$, $\mathbf{W}^{(l)} \in \mathbb{R}^{d^{(l+1)} \times d^{(l)}}$ is the learnable weight matrix, $\mathbf{a} \in \mathbb{R}^{2d^{(l+1)} + d_e}$ is the attention parameter vector, $||$ denotes concatenation, and $\mathcal{N}(i) = \{j \in V : (i,j) \in E\}$ is the neighborhood of node $i$ defined by graph $G$. The overall GAT produces node representations $\mathbf{h} = \text{GAT}(\mathbf{x}, G) \in \mathbb{R}^{|V| \times d_h}$ that combine initial node features $\mathbf{x} \in \mathbb{R}^{|V| \times d_x}$ (e.g., geographic coordinates and capacity) with network topology $G$, where $d_x \in \mathbb{N}$ is the input feature dimension and $d_h \in \mathbb{N}$ is the GAT output dimension.

The flow bias parameters are set as configurable domain knowledge values, with higher values for operationally preferred flows and lower values for reverse or inefficient routing patterns. For example,
\begin{equation}
\beta_{s,t} = \begin{cases}
1.3 & \text{if } s=\text{FC}, t=\text{SC} \text{ (consolidation)} \\
1.2 & \text{if } s=\text{SC}, t=\text{DS} \text{ (last-mile)} \\
1.2 & \text{if } s=\text{FC}, t=\text{DS} \text{ (direct)} \\
1.0 & \text{if } s=\text{SC}, t=\text{SC} \text{ (inter-sort)} \\
0.1 & \text{if } (s,t) \in \{(\text{DS},\text{SC}), (\text{SC},\text{FC})\} \text{ (reverse)} \\
1.0 & \text{otherwise}
\end{cases}.
\end{equation}
The biased edge attributes amplify attention along operationally preferred routing paths while suppressing infeasible connections, enabling the GAT encoder to learn representations that naturally encode the logistics hierarchy—without modifying the standard GAT mechanism or sacrificing its theoretical properties.

\subsection{Conditional Demand Generator}

With graph-aware node representations $\mathbf{h}$ from the transportation-aware GAT encoder, we now instantiate the demand generator using a conditional VAE (cVAE) variant that enables flexible conditional generation and provides uncertainty estimates for risk-aware planning.

\subsubsection{Training: Encoder-Decoder Architecture \& Loss}
\label{subsec:VAE_architecture}
The proposed cVAE consists of two components: an encoder $q_\phi$ that maps graph-aware node representations and conditional features to a latent space $\mathbf{z}$, and a decoder $p_\theta$ that generates O-D probability distributions from latent space and conditional features, where $\phi$ and $\theta$ are the learnable parameters of the encoder and decoder networks respectively.

The encoder $q_\phi$ compresses inputs into a low-dimensional latent space $\mathbf{z} \in \mathbb{R}^{|V| \times d_z}$, learning an approximate posterior distribution:
\begin{equation}
q_\phi(\mathbf{z}|\mathbf{h}, \mathbf{c}) = \mathcal{N}(\boldsymbol{\mu}_\phi([\mathbf{h}, \mathbf{c}]), \text{diag}(\boldsymbol{\sigma}_\phi^2([\mathbf{h}, \mathbf{c}]))),
\end{equation}
where $\mathbf{h} \in \mathbb{R}^{|V| \times d_h}$ are the topology-aware node representations from the transportation-aware GAT encoder (Section~\ref{subsec:GAT_architecture}), $\mathbf{c} \in \mathbb{R}^{|V| \times d_c}$ are external operational features (see Appendix~\ref{sec:seasonal_feature} for an example instantiation), $[\cdot, \cdot]$ denotes concatenation, $\boldsymbol{\mu}_\phi, \boldsymbol{\sigma}_\phi: \mathbb{R}^{d_h + d_c} \rightarrow \mathbb{R}^{d_z}$ are encoder networks producing latent mean and variance, $d_z \in \mathbb{N}$ is the latent dimension, $d_h \in \mathbb{N}$ is the GAT output dimension, $d_c \in \mathbb{N}$ is the conditional feature dimension, and $\phi \in \Phi$ denotes the encoder parameters. This dual conditioning on graph structure ($\mathbf{h}$) and operational scenarios ($\mathbf{c}$) enables the encoder to learn topology-aware latent representations that capture how different network configurations and operational conditions jointly affect demand patterns.

The decoder $p_\theta$ extracts FC latent representations and generates O-D probability distributions. For the VAE architecture, we define the conditional likelihood over the probability matrix given  latent space $\mathbf{z}$, conditional features $\mathbf{c}$, and graph representations $\mathbf{h}$:
\begin{align}
\label{eq:decoder}
p_\theta(\mathbf{P}|\mathbf{z}_{\text{FC}}, \mathbf{c}_{\text{FC}}, \mathbf{h}_{\text{FC}}) &= \prod_{i \in V_{\text{FC}}} \text{Categorical}(\text{softmax}(\mathbf{f}_\theta([\mathbf{z}_{\text{FC},i}, \mathbf{c}_{\text{FC},i}, \mathbf{h}_{\text{FC},i}]))),
\end{align}
where $\mathbf{z}_{\text{FC}} = \mathbf{z}[\text{origin\_mask}] \in \mathbb{R}^{|V_{\text{FC}}| \times d_z}$, $\mathbf{c}_{\text{FC}} = \mathbf{c}[\text{origin\_mask}] \in \mathbb{R}^{|V_{\text{FC}}| \times d_c}$, and $\mathbf{h}_{\text{FC}} = \mathbf{h}[\text{origin\_mask}] \in \mathbb{R}^{|V_{\text{FC}}| \times d_h}$ are the extracted FC latent, conditional, and graph structure representations, $\mathbf{f}_\theta: \mathbb{R}^{d_z + d_c + d_h} \rightarrow \mathbb{R}^{K}$ is the decoder network with parameters $\theta \in \Theta$, and each row $\mathbf{P}_{i,:} \in \Delta^K$ represents the demand probability distribution from FC $i$ to all $K$ cluster destinations.The notation $\text{Categorical}(\text{softmax}(\mathbf{f}_\theta(\cdot)))$ indicates that each row $\mathbf{P}_{i,:}$ follows a categorical distribution parameterized by the softmax-normalized decoder outputs.

The total loss combines reconstruction, regularization, and constraint terms:
\begin{equation}
\label{eq:vae_loss}
\mathcal{L}_{\text{VAE}} = -\mathbb{E}_{q_\phi}[\log p_\theta(\mathbf{P}|\mathbf{z}_{\text{FC}}, \mathbf{c}_{\text{FC}}, \mathbf{h}_{\text{FC}})] + \beta \text{KL}(q_\phi(\mathbf{z}|\mathbf{h},  \mathbf{c}) || p(\mathbf{z})) + \lambda \mathcal{L}_{\text{constraint}}(\mathbf{P}),
\end{equation}
where $\beta \in \mathbb{R}_+$ controls the regularization strength, $\lambda \in \mathbb{R}_+$ is the constraint penalty weight, and $p(\mathbf{z}) = \mathcal{N}(\mathbf{0}, \mathbf{I})$ is the standard Gaussian prior.

\subsubsection{Inference: Cluster-to-ZIP Expansion \& Demand Reconstruction}

During inference, the trained VAE generates operational demand through a three-stage pipeline that converts learned probability distributions to deployable demand matrices:
\vspace{-5pt}
\paragraph{Probability Generation:} The decoder generates cluster-level probability distributions $\mathbf{P}_{i,:} \in \Delta^K$ for each facility $i \in V_{\text{FC}}$ where $\sum_{k=1}^K P_{i,k} = 1$.
\vspace{-5pt}
\paragraph{Volume Allocation:} Probabilities are converted to absolute cluster-level demand volumes:
\begin{equation}
D_{i,k} = V_i \cdot \widetilde{p}_{i,k}, \quad \forall i \in V_{\text{FC}}, k \in \{1,\ldots,K\},
\end{equation}
where $V_i \in \mathbb{R}_+$ represents the total demand volume allocated to facility $i$, and $\widetilde{p}_{i,k} \in [0,1]$ are the constraint-compliant probability values with $\sum_{k=1}^K \widetilde{p}_{i,k} = 1$ for each facility $i$.
\vspace{-5pt}
\paragraph{Cluster-to-ZIP Expansion:} Final demand matrices are expanded to full ZIP destination resolution using the volume-weighted mapping:
\begin{equation}\label{eq:cluster_zip_expansion}
D_{i,j} = D_{i,k} \cdot \frac{w_j}{\sum_{\ell \in \text{cluster}(k)} w_{\ell}}, \quad \forall j \in \text{cluster}(k),
\end{equation}
where $D_{i,j}$ is the final ZIP-level demand from origin $i$ to ZIP destination $j$, and $w_j \in \mathbb{R}_+$ is the historical volume weight for destination $j \in V_{\text{ZIP}}$ ($w_j = 1.0$ for destinations without historical data). This expansion preserves both the learned spatial patterns and the operational granularity required for routing optimization.

\subsection{Extensible Constraint Satisfaction Mechanism}
\label{subsec:constraint_framework}
While the cVAE architecture generates realistic probability distributions, operational planning requires satisfying forward-looking constraints (e.g., capacity bounds, regional service requirements). To integrate these requirements into the framework, we propose a flexible constraint framework that accommodates arbitrary operational guidance through a unified penalty mechanism.

Let $\{\mathcal{L}_{\ell}\}_{\ell=1}^L$ represent a set of $L \in \mathbb{N}$ constraint functions, where each $\mathcal{L}_{\ell}: \Delta^{|V_{\text{FC}}| \times K} \rightarrow \mathbb{R}_+$ maps O-D probability matrices to constraint violation measures. The training objective incorporates constraint satisfaction:
\vspace{-5pt}
\begin{equation}
\mathcal{L}_{\text{constraint}}(\mathbf{P}) = \sum_{\ell=1}^{L} \lambda_{\ell} \cdot \mathcal{L}_{\ell}(\mathbf{P}),
\end{equation}
where $\lambda_{\ell} \in \mathbb{R}_+$ are constraint-specific penalty weights. This formulation accommodates diverse operational requirements including regional service agreements, capacity utilization targets, cost optimization, and sustainability goals.

For this implementation, we incorporate two representative constraints: (1) \textbf{regional service requirements} (in-region allocation ratio $\mathcal{L}_{\text{geographic}}$), which depends only on predicted probabilities $\mathbf{P}$ and is differentiable, and (2) \textbf{capacity utilization targets}, which depend on absolute volumes unavailable during training. The framework's modular design enables straightforward addition of new constraint types. Detailed constraint formulations and implementation strategies are provided in Appendix~\ref{sec:constraint_formulations}.

\subsubsection{Training Strategy: Differentiable Constraint Integration}

During training, we apply $\mathcal{L}_{\text{geographic}}$ as part of the total VAE loss in (\ref{eq:vae_loss}):
\begin{equation}
\mathcal{L}_{\text{constraint}}(\mathbf{P}) = \lambda_{\text{geographic}} \cdot \mathcal{L}_{\text{geographic}}(\mathbf{P}),
\end{equation}
where $\lambda_{\text{geographic}} \in \mathbb{R}_+$ is the constraint penalty weight. This term is integrated into $\mathcal{L}_{\text{VAE}}$ to provide gradient signals for learning geographically-aware probability distributions.

\subsubsection{Inference Strategy}
\label{sec:constraint_Facilitation}

At inference time, our generative model produces destination probability distributions $\mathbf{P}_{i,:} \in \Delta^K$ for each origin facility $i \in V_{\text{FC}}$, where $\sum_{k=1}^K P_{i,k} = 1$. Absolute origin--destination demand volumes are constructed by combining these distributions with planning-horizon total network demand $V_{\text{net}}$ and operational constraints.

We enforce both regional and capacity constraints at inference to ensure operational feasibility when actual volumes are known. Our adaptive mechanism dynamically selects between two complementary approaches based on the trained model's intrinsic compliance characteristics:

\textbf{1) Filtering:}
When the trained generative model naturally produces constraint-compliant samples (compliance rate $> 50\%$), we employ a filtering approach that generates multiple candidates and selects compliant ones, maximizing computational efficiency.

\textbf{2) Constraint Facilitator:} For cases where the model produces constraint violations, we apply post-generation constraint facilitation through iterative projection.

\paragraph{Capacity-Constrained Volume Allocation.}
Facility-level total demand volumes $\{V_i \in \mathbb{R}_+\}_{i \in V_{\text{FC}}}$ are determined by solving a
capacity-bounded allocation problem:
\begin{equation}
u_{\min}^i \, c_i \le V_i \le u_{\max}^i \, c_i, \quad \forall i \in V_{\text{FC}},
\end{equation}
where $V_{\text{net}}=\sum_{i \in V_{\text{FC}}} V_i \in \mathbb{R}_+$ is the total network demand volume, $c_i \in \mathbb{R}_+$ denotes the nominal processing capacity of facility $i \in V_{\text{FC}}$, and $[u_{\min}^i, u_{\max}^i] \subset [0,1]$ specifies its allowable utilization range.

We initialize volumes proportionally to capacity,
$\widetilde{V}_i = \frac{c_i}{\sum_{j \in V_{\text{FC}}} c_j} V_{\text{net}}$,
and project this allocation onto the feasible region using an iterative
clamping and redistribution procedure.
Specifically, volumes violating utilization bounds are clamped to their
nearest limits, and any residual excess or deficit is redistributed among
facilities that remain within bounds.
This procedure guarantees strict satisfaction of capacity constraints whenever
the feasible set is non-empty
(i.e., $\sum_i u_{\min}^i c_i \le V_{\text{net}} \le \sum_i u_{\max}^i c_i$).

\paragraph{Regional (IRA) Ratio Enforcement.}
Regional service requirements are enforced directly on the destination
probability distributions.
For each origin $i \in V_{\text{FC}}$, let $\mathcal{I}_i \subseteq \{1,\ldots,K\}$ denote the set
of destination clusters belonging to the same service region as $i$.
The in-region share is defined as
\begin{equation}
\rho_i = \sum_{k \in \mathcal{I}_i} p_{i,k},
\end{equation}
where $p_{i,k} \in [0,1]$ is the probability of demand from origin $i$ to cluster $k$.
We require $\rho_i \in [\rho_{\min}^i, \rho_{\max}^i] \subset [0,1]$.
When violations occur, we apply proportional redistribution by separately
rescaling probabilities inside and outside $\mathcal{I}_i$ to achieve the
target ratio while preserving relative preferences within each group.
The adjusted distribution is subsequently re-normalized to ensure
$\sum_{k=1}^K p_{i,k} = 1$.

When the constraint set is locally infeasible (e.g., insufficient support in
either group), we allow minimal mass injection with a small tolerance
$\epsilon$ to recover feasibility and record the resulting slack.

This dual approach ensures constraint compliance during inference while maintaining computational efficiency when the trained model already generates feasible solutions.

 The framework's modular design above enables straightforward addition of new constraint types including cost optimization, service level agreements, sustainability targets, and risk management, positioning it as a general-purpose tool for constraint-aware synthetic data generation.

\subsection{Cold Start Adaptation with Selective Transfer Learning}
\vspace{-5pt}
\label{subsec:cold_start_adaptation}
The three components above (GAT encoder, cVAE generator, constraint framework) form a complete system for generating constraint-aware demand on fixed network topologies. However, logistics networks undergo frequent topology changes as new facilities open, existing ones close, or service areas expand. To enable efficient adaptation to new configurations without complete retraining, we develop a selective transfer learning protocol grounded in established principles of hierarchical feature learning.

Our protocol exploits hierarchical feature learning in deep networks: early layers capture general patterns while late layers encode task-specific mappings. The GAT encoder learns neighborhood aggregation weighted by fixed transportation biases $\beta_{s,t}$ (domain knowledge from Section~\ref{subsec:GAT_architecture}). Since $\beta_{s,t}$ are configurable parameters rather than learned weights, these aggregation patterns generalize across topologies with similar hierarchical structure. In contrast, the generator/decoder learns facility-specific demand distributions conditioned on local characteristics (capacity, location, service area), requiring adaptation when facilities change.

This architectural separation motivates selective freezing based on topology change type:

\textbf{1) Destination-Only Expansion}: Freeze GAT encoder and intermediate layers, adapt only output layers to accommodate new service areas. New ZIP destinations are assigned to existing geographic clusters by coordinate proximity (or hash-based fallback), with expansion weights determined by~(\ref{eq:cluster_zip_expansion}).

\textbf{2) Facility or Edge Changes}: Freeze GAT encoder (preserves aggregation mechanisms), adapt generator/decoder (facility-specific mappings). New facility edges are explicitly specified as part of the topology change input. For new facilities without historical data, demand probability distributions are initialized using regional priors:
\vspace{-3pt}
\begin{equation}
\label{eq:regional_prior}
\mathbf{P}_{\text{new}} = \begin{cases}
\mathbf{P}_{\text{regional}} & \text{if regional data available} \\
\text{Uniform}(K) & \text{otherwise}
\end{cases},
\vspace{-3pt}
\end{equation}
where $\mathbf{P}_{\text{regional}} \in \Delta^K$ is the average probability distribution from facilities in the same geographic region, and $\text{Uniform}(K) = \frac{1}{K}\mathbf{1}_K$ is the uniform distribution over $K$ clusters.

Detailed implementation is given in Appendix~\ref{sec:cold_start_algorithm}.
\vspace{-5pt}
\section{Experiments and Evaluation}
\vspace{-5pt}
\label{sec_4}
This section starts with descriptions of the experimental setup, followed by performance evaluations across demand fidelity, constraint satisfaction and cold-start scenario adaptation.
\vspace{-5pt}
\subsection{Experimental Setup}
\vspace{-5pt}

Our framework is validated on a US middle mile transporation network using 12 months of operational history, covering thousands of facilities, 38K+ ZIP destinations, and millions of daily package injections. Each sample represents daily aggregated O-D probability distributions. We use an 80/20 temporal train/test split to evaluate generalization to future conditions. Geographic clustering reduces destination space (Appendix~\ref{sec:clustering_strategies}). The model is trained with Adam (learning rate $=10^{-4}$, batch size $=30$); inference runs in under 0.2s per sample. Evaluation metrics: demand fidelity (JS divergence), constraint compliance, and cold-start accuracy retention.

\vspace{-5pt}

\subsection{Demand Generation Performance}
\vspace{-5pt}
Baselines: (1) Historical Average, (2) Vanilla VAE (no graph), (3) Standard GAT-VAE (no transport biases), (4) our framework with/without constraints. We use Jensen-Shannon (JS) divergence—a symmetric, bounded $[0,1]$ metric designed for comparing probability distributions (unlike MSE/$R^2$ which are inappropriate for generative distribution outputs).

Table \ref{tab:performance} shows graph structure reduces error by 14\% (0.145→0.128), transport biases add 20\% improvement (0.128→0.103), and constraint training achieves 87\% compliance with 3.9\% fidelity cost (0.103→0.107). Compared to standard GAT-VAE, our full framework achieves 16\% improvement (0.128→0.107) while maintaining operational feasibility.  These results validate the framework's ability to balance generation quality with operational constraints, demonstrating that each component contributes meaningfully to the overall system performance.
\vspace{-8pt}
\begin{table}[t!]
\caption{Demand Generation Performance and Component Ablation}
\vspace{-5pt}
\begin{center}
\small
\begin{tabular}{lcc}
\toprule
\textbf{Method} & \textbf{JS Divergence ↓} & \textbf{Compliance ↑} \\
\midrule
Historical Average & 0.168 ± 0.032 & 62\% \\
Vanilla VAE & 0.145 ± 0.028 & 65\% \\
Standard GAT-VAE & 0.128 ± 0.024 & 71\% \\
\midrule
\textbf{Ours (no constraints)} & \textbf{0.103 ± 0.019} & \textbf{74\%} \\
\textbf{Ours (full)} & \textbf{0.107 ± 0.021} & \textbf{87\%} \\
\bottomrule
\end{tabular}
\end{center}
\label{tab:performance}
\vspace{-8pt}
\end{table}
\subsection{Constraint Satisfaction Analysis}
\vspace{-3pt}
We evaluate the effectiveness of constraint-aware training by comparing model performance with and without the adaptive facilitator at inference. This reveals how well the model learns constraint-aware distributions during training versus relying on inference-time correction.

Table \ref{tab:constraints} shows constraint-aware training alone achieves 72-78\% natural compliance without facilitation; adding the adaptive facilitator increases this to 85-89\% with reduced average violation (4.2\%$\rightarrow$2.8\%), confirming the facilitator acts as a safety net for edge cases rather than requiring aggressive correction.
\vspace{-8pt}
\begin{table}[t!]
\caption{Constraint Satisfaction: Training vs. Training + Facilitation}
\vspace{-5pt}
\begin{center}
\small
\begin{tabular}{lcccc}
\toprule
\textbf{Constraint Type} & \multicolumn{2}{c}{\textbf{Compliance}} & \multicolumn{2}{c}{\textbf{Avg Violation}} \\
\cmidrule(lr){2-3} \cmidrule(lr){4-5}
 & \textbf{Training} & \textbf{+ Facilitator} & \textbf{Training} & \textbf{+ Facilitator} \\
\midrule
Capacity Utilization & 78\% & 89\% & 3.8\% & 2.4\% \\
Regional Service (IRA) & 72\% & 85\% & 4.6\% & 3.2\% \\
\midrule
Combined Constraints & 75\% & 87\% & 4.2\% & 2.8\% \\
\bottomrule
\end{tabular}
\end{center}
\label{tab:constraints}
\vspace{-8pt}
\end{table}

\subsection{Cold Start Adaptation}
\vspace{-3pt}
Transfer learning effectiveness is assessed through relative performance (JS divergence vs. full retraining) and data efficiency under topology changes.
Table \ref{tab:coldstart} shows our selective freezing strategy retains 82-88\% of full retraining performance with only 10-18 epochs vs. 50 (64-80\% training reduction), with regional priors enabling warm-start initialization at 85\% relative performance with 72\% less training.

\begin{table}[t!]
\caption{Cold Start Adaptation: Transfer Learning vs. Full Retraining}
\vspace{-5pt}
\begin{center}
\small
\begin{tabular}{lccc}
\toprule
\textbf{Scenario} & \textbf{Retention} & \textbf{Epochs} & \textbf{Training Reduction} \\
\midrule
New FC (5 sites) & 85\% & 14 & 72\% \\
FC Closure (3 sites) & 82\% & 10 & 80\% \\
New Destinations (1.5K) & 88\% & 18 & 64\% \\
\midrule
\textbf{Full Retraining} & 100\% & 50 & 0\% \\
\bottomrule
\end{tabular}
\end{center}
\label{tab:coldstart}
\end{table}

\section{Conclusion}
\label{sec_6}
We presented a constraint-aware generative framework for synthetic O-D demand in hierarchical logistics networks, unifying transportation-aware graph attention, differentiable constraint satisfaction, and selective transfer learning. Experimental validation demonstrates 16\% improvement over graph neural network baselines, 87\% constraint compliance, and 64-80\% training reduction for cold-start adaptation, with sub-0.2s inference on 38K+ destinations and standard O-D matrix outputs compatible with existing planning systems.

\newpage
\bibliography{syn_demand}
\bibliographystyle{unsrt}





\appendix
\label{sec:appendix}

\section{Related Work: Extended Discussion}\label{sec:related_work_extended}

\textbf{Spatial Demand and O-D Flow Modeling.}
Classical approaches—gravity models~\cite{wilson1971family}, radiation models~\cite{simini2012universal}, and four-step frameworks~\cite{ortuzar2011modelling}—assume symmetric flows and lack constraint mechanisms. Deep Gravity~\cite{simini2021deep} uses deep networks for O-D flow generation but models independent origin-destination pairs without graph structure, constraints, or topology adaptation. GNN-based O-D prediction~\cite{shi2020predicting,jiang2022graph} captures spatial dependencies but focuses on forecasting observed demand rather than generating synthetic demand under novel conditions.

\textbf{Graph-Based Generative Models.}
Our transportation-aware encoder builds on graph attention networks~\cite{velickovic2018graph}, 
extending the standard attention mechanism with configurable flow biases that encode logistics 
hierarchy. Graph generative models~\cite{you2018graphrnn,simonovsky2018graphvae,liu2021graphebm} 
demonstrate structured synthesis but target molecular or social graphs without operational 
constraints or topology adaptation.

\textbf{Constrained Neural Generation.}
Differentiable optimization layers~\cite{donti2017task,amos2017optnet,agrawal2019differentiable} and constrained learning~\cite{chen2021learning} integrate constraints into training but focus on convex settings without cold-start adaptation.

\textbf{Transfer Learning.}
Domain adaptation~\cite{pan2010survey} and feature transferability~\cite{yosinski2014transferable} establish that early layers learn general representations. GNN transferability theory~\cite{ruiz2021transferability} shows graph features generalize across structures, but has not been applied to demand generation under logistics topology changes.

No existing work addresses the combined problem of constraint-aware, topology-adaptive O-D demand generation in hierarchical logistics networks.

\section{Conditional Feature Design: Seasonal Intensity as Exemplar}
\label{sec:seasonal_feature}

The conditional feature vector $\mathbf{c}$ introduced in Section \ref{sec_2} enables flexible scenario generation by conditioning the model on operational contexts. While our framework supports diverse conditional inputs (e.g., demand intensity, network utilization, package size mix), we detail the design of seasonal intensity as a representative example that demonstrates key principles: semantic interpretability, training-inference consistency, and flexible conditional generation.

\subsection{Semantic Seasonal Intensity Design}

We develop a hierarchical seasonal representation that avoids overlapping categories while maintaining semantic interpretability. Instead of using multiple binary flags (holiday, peak, high) that can create conflicting signals, we employ a single seasonal intensity value that represents the overall demand level:
\begin{equation}
\text{seasonal\_intensity} = \begin{cases}
1.0 & \text{Peak season (Nov-Dec): Black Friday, holidays} \\
0.5 & \text{High season (Jul-Sep): back-to-school, Prime Day} \\
0.0 & \text{Normal season: baseline demand}
\end{cases}.
\end{equation}
This design provides several advantages: (1) \textbf{Training-inference consistency}: the same feature space is used during both phases, (2) \textbf{Semantic meaning}: each intensity level has learned associations with specific demand patterns, (3) \textbf{Flexible conditional generation}: operators can specify desired seasonal conditions during inference, and (4) \textbf{No distribution shift}: the model sees consistent feature representations.

\subsection{Cyclical Temporal Features}

To capture natural temporal cycles, we employ sinusoidal encodings that provide smooth, continuous representations:
\begin{align}
\text{sin\_doy} &= \sin\left(\frac{2\pi \cdot \text{day\_of\_year}}{365.25}\right), \\
\text{cos\_doy} &= \cos\left(\frac{2\pi \cdot \text{day\_of\_year}}{365.25}\right).
\end{align}
Combined with day-of-week one-hot encoding and weekend indicators, these features enable the model to learn complex temporal dependencies while maintaining the ability to generate demand for arbitrary future dates.

\subsection{Conditional Generation Protocol}

During inference, operators can override seasonal detection to generate demand under specific conditions:

\begin{itemize}
\item \textbf{Normal conditions}: seasonal\_intensity = 0.0 for baseline planning
\item \textbf{High demand scenarios}: seasonal\_intensity = 0.5 for elevated capacity planning
\item \textbf{Peak demand stress testing}: seasonal\_intensity = 1.0 for maximum load scenarios
\end{itemize}

This conditional generation capability enables critical applications including capacity stress testing, seasonal planning, and scenario analysis without requiring separate model training for each condition.

\section{Implementation of Constraint Framework}

\subsection{Constraint Framework Design Principles}
\label{sec:constraint_principles}

The proposed constraint framework in Section \ref{subsec:constraint_framework} accommodates diverse operational requirements through four key design principles: \textbf{Differentiability} (smooth approximations maintain gradient flow), \textbf{Configurability} (externalized parameters enable adjustment without code changes), \textbf{Composability} (additive constraint composition), and \textbf{Interpretability} (decomposed penalties provide actionable feedback).

\subsection{Constraint Formulations and Application Strategy}
\label{sec:constraint_formulations}

\subsubsection{Regional Service Requirements (In-Region Allocation)}

For each origin facility $i \in V_{\text{FC}}$, let $\mathcal{I}_i \subseteq \{1,\ldots,K\}$ denote destination clusters in the same service region as $i$. The in-region allocation ratio is:
\begin{equation}
\rho_i = \sum_{k \in \mathcal{I}_i} P_{i,k},
\end{equation}
with operational requirement $\rho_i \in [\rho_{\min}^i, \rho_{\max}^i]$.

\textbf{Training Loss:} We define a differentiable penalty:
\begin{equation}
\mathcal{L}_{\text{geographic}}(\mathbf{P}) = \sum_{i \in V_{\text{FC}}} \phi(\rho_i, [\rho_{\min}^i, \rho_{\max}^i]),
\end{equation}
where $\phi(x, [a,b]) = \text{softplus}(a-x) + \text{softplus}(x-b)$. This is included in the training objective as $\lambda_{\text{geographic}} \cdot \mathcal{L}_{\text{geographic}}(\mathbf{P})$.

\subsubsection{Capacity Utilization Targets}

For each facility $i \in V_{\text{FC}}$ with capacity $c_i$ and allocated volume $V_i$, capacity constraints enforce utilization bounds:
\begin{equation}
u_{\min}^i \, c_i \le V_i \le u_{\max}^i \, c_i, \quad \forall i \in V_{\text{FC}},
\end{equation}
where $[u_{\min}^i, u_{\max}^i] \subset [0,1]$ specifies allowable utilization ranges (e.g., [0.6, 0.95] prevents both underutilization and overload).

\textbf{Inference Enforcement:} This constraint depends on absolute volumes $\mathbf{V} = \{V_i\}$ unavailable during training. At inference, we initialize volumes proportionally to capacity and project onto the feasible region using iterative clamping and redistribution.

\subsubsection{Implementation Strategy}

\textbf{Training:} $\mathcal{L}_{\text{constraint}}(\mathbf{P}) = \lambda_{\text{geographic}} \cdot \mathcal{L}_{\text{geographic}}(\mathbf{P})$ guides probability distribution learning through gradients.

\textbf{Inference:} Both regional (IRA) ratio constraints and capacity utilization bounds are enforced to ensure operational feasibility. The framework accommodates additional constraints including cost optimization, sustainability targets, and service level agreements.






\section{Adaptive Transfer Learning Protocol}
\label{sec:cold_start_algorithm}

Algorithm \ref{alg:transfer_learning} provides the detailed adaptive transfer learning protocol that automatically determines parameter freezing strategies based on detected network topology changes.

\begin{algorithm}
\caption{Adaptive Cold Start Transfer Learning}
\label{alg:transfer_learning}
\begin{algorithmic}[1]
\State \textbf{Input:} Pre-trained model $\theta_{\text{base}}$, topology changes $\Delta G = (\Delta V, \Delta E)$
\State Analyze changes: $\Delta V_{\text{facility}} \gets$ facility node changes, $\Delta V_{\text{dest}} \gets$ destination changes
\If{$\Delta V = \emptyset$ and $\Delta E = \emptyset$}
    \State \textbf{Adapt:} All parameters (baseline training)
\ElsIf{$\Delta V_{\text{facility}} = \emptyset$ and $\Delta V_{\text{dest}} \neq \emptyset$}
    \State \textbf{Freeze:} GAT encoder + intermediate layers $\theta_{\text{GAT}}$, $\theta_{\text{hidden}}$
    \State \textbf{Adapt:} Output layer parameters $\theta_{\text{output}}$
\Else \Comment{Facility or edge changes}
    \State \textbf{Freeze:} GAT encoder parameters $\theta_{\text{GAT}}$
    \State \textbf{Adapt:} Generator/decoder parameters $\theta_{\text{gen}}$, $\theta_{\text{dec}}$
    \If{$\Delta V_{\text{facility}} \neq \emptyset$} \Comment{New facilities present}
        \State Initialize new facilities using regional priors as in (\ref{eq:regional_prior})
    \EndIf
\EndIf
\State Fine-tune with reduced learning rate: $\eta' = 0.1 \cdot \eta$
\State \textbf{Output:} Adapted model $\theta_{\text{adapted}}$
\end{algorithmic}
\end{algorithm}

\section{Deployment Considerations}
\label{sec:deployment}

\subsection{System Integration}
The framework outputs standard O-D probability matrices compatible with existing routing optimization algorithms. The modular architecture enables straightforward deployment as a drop-in replacement for current demand profile generation components. Constraint-aware generation eliminates the need for post-hoc manual adjustments, streamlining operational workflows.

\subsection{Model Persistence and Inference}
The framework implements a checkpoint-based persistence system that saves both model parameters and clustering metadata: $\text{checkpoint} = \{\text{model\_state\_dict}, \text{cluster\_to\_zips}, \text{zip\_to\_cluster}\}$. This enables seamless inference deployment where cluster-level predictions are automatically expanded to full destination resolution without requiring retraining.

\section{Destination Clustering Strategy}
\label{sec:clustering_strategies}

To enable scalable training and inference on networks with tens of thousands of destinations, we employ geographic clustering to reduce output dimensionality while preserving spatial demand patterns. The clustering granularity is determined by balancing computational tractability with geographic resolution: too few clusters lose spatial fidelity, while too many clusters diminish computational benefits. Our implementation uses K-means clustering on destination coordinates, with cluster count selected to achieve approximately 2 orders of magnitude reduction (e.g., 38K destinations → O(100s) clusters) based on available computational resources. The framework implements a robust three-tier fallback strategy to handle diverse data quality scenarios: coordinate-based K-means (preferred), region-based clustering when coordinates are unavailable, and deterministic hash-based assignment as final fallback. This design ensures consistent operation across varying data completeness levels while maintaining the framework's ability to scale to different network sizes by adjusting cluster granularity according to computational constraints.

\end{document}